\documentclass{nature}
\usepackage{amsmath,amssymb,amsfonts,bm,mathtools}
\usepackage{graphicx}
\let\CERTincludegraphics\includegraphics
\usepackage{booktabs}
\usepackage{array}
\usepackage{xcolor}
\usepackage{enumitem}
\usepackage[T1]{fontenc}
\usepackage{microtype}
\usepackage{caption}
\DeclareCaptionLabelSeparator{nature}{: }
\usepackage{pdfpages}
\usepackage{hyperref}

\makeatletter
\AtBeginDocument{\let\includegraphics\CERTincludegraphics}
\renewenvironment*{figure}[1][tbp]{\@float{figure}[#1]\small}{\end@float}
\renewenvironment*{table}[1][tbp]{\@float{table}[#1]\small}{\end@float}
\makeatother

\title{Detecting False Self-Consistency in Density-Functional Calculations}
\author{Mohsen Sotoudeh$^{1,2}$}

\newcommand{\vth}{\bm{\theta}}
\newcommand{\vout}{\bm{v}_{\mathrm{out}}}

\newcommand{\SI}{Supplementary Information}
\newcommand{\runin}[1]{\subsection{#1}}
\newcommand{\reshead}[1]{\par\medskip\noindent{\bfseries #1}\par\nopagebreak\smallskip}

\begin{document}
\maketitle

\begin{affiliations}
 \item Helmholtz Institute Ulm (HIU), Electrochemical Energy Storage, 89081 Ulm, Germany
 \item Karlsruhe Institute of Technology (KIT), P.O. Box 3640, 76021 Karlsruhe, Germany
\end{affiliations}

\begin{abstract}
\noindent
Density-functional calculations feed materials databases and
machine-learning models, and are declared converged
when energy changes and residuals fall below thresholds. We show that
these criteria, and even the agreement of independent optimizers, can
accept a state from which the energy still descends or whose
Hamiltonian is inconsistent with the energy. A variational formulation of
self-consistency in the space of Kohn-Sham Hamiltonians yields a
descent probe that exposes lower states accessible from a returned
solution, and a frozen-orbital Janak test detects such an
inconsistency. Both apply to any code without modification. In VASP, a UO$_2$ PBE+$U$ calculation converges
1.46~eV above a lower state, and CoO and FeO converge as metals
0.8--0.9~eV above insulating states. In a census of 339 converged
calculations on tungsten-free Materials Project structures, recomputed
with a modified static protocol, 57 (16.8\%) show energy decreases above the threshold, 49 without an initial energy rise exceeding the tolerance.
\end{abstract}

\section*{Introduction}
Density-functional calculations~\cite{hohenberg64_pr136_B864,
  kohn65_pr140_A1133} are now routinely carried out in automated
workflows, which produce materials databases and the training data of
machine-learned models~\cite{jain13_aplmater1_011002,
  curtarolo12_cms58_218,saal13_jom65_1501,huber20_scidata7_300}. The
results of different codes agree at the level of
meV~\cite{lejaeghere16_science351_aad3000} in comparisons with matched
settings, but whether an individual calculation has reached an
acceptable solution is rarely tested. A
calculation is declared converged when successive energies agree and the
residual of the density or potential is small. Neither criterion is
sufficient: a stalled or oscillating iteration can fulfill both, and so
can an unstable self-consistent solution or the self-consistent
solution of a Hamiltonian that is not the derivative of the reported
energy. The common practical safeguard, agreement between different
algorithms, fails as well when the algorithms share the same
Hamiltonian. For high-throughput data the consequence is not only an
energy error: a falsely converged state can carry the wrong magnetic
moment or the wrong band gap, and even the wrong sign of the gap, into
every quantity and every model derived from it.

Three properties are at stake and must be distinguished. A state is
self-consistent if its density and Hamiltonian satisfy the fixed-point
condition of the Kohn-Sham equations. It is variationally stable if no
descent of the implemented free energy is accessible from it within the
tested variational space. A self-consistent state can be a saddle
point. The Hamiltonian is consistent with the energy if it is the
derivative of the implemented energy functional. Conventional stopping
criteria can wrongly suggest any of the three. We use false
self-consistency as an umbrella term for a calculation that is reported
as converged but fails at least one of them, not only for states that
violate the fixed-point equations.

Such failures occur in practice. A PBE+$U$ calculation of UO$_2$ in VASP
that reports convergence lies 1.46~eV above the state reached by a
descent from its own orbitals (Fig.~\ref{fig:uo2}). Under a static
protocol derived from that of the Materials
Project~\cite{jain13_aplmater1_011002}, CoO
converges to a metal 0.79~eV above an insulating state with a gap of
1.40~eV. With a deliberately reintroduced Hamiltonian inconsistency,
five independent optimizers agree on a state 7.5~meV too high
[Fig.~\ref{fig:certify}c].

Here we show that a variational formulation of the self-consistency
problem in the space of Hamiltonians, together with Janak's theorem,
provides two tests that can falsify such a result. They do not depend
on the optimizer that produced the state, they distinguish a state from
which a descent remains from a Hamiltonian that is inconsistent with
the energy, and they apply to other codes without changes. We
demonstrate them on controlled failures in our own code, on $f$-electron
oxides in VASP, on 22 targeted magnetic oxides computed with a database
protocol, and on a census of 366 magnetic oxide structures from the
Materials Project, which measures how often a calculation that reports
convergence under this protocol retains a variational descent beyond
its tolerance. We also state precisely what the tests cannot establish,
and we give a recipe by which they can be added to an automated
workflow, either as an early warning or as a full correction.

\section*{Results}
\reshead{Free energy as a function of the Hamiltonian}
In the projector augmented-wave (PAW) method~\cite{bloechl94_prb50_17953},
the Kohn-Sham Hamiltonian
$\tilde H[\vth]=\hat T+\hat H_0+\sum_\alpha\theta_\alpha\hat\rho_\alpha$,
where $\hat T$ is the kinetic-energy operator and $\hat H_0$ contains
the parameter-independent terms, is linear in the parameters
$\vth=(\tilde v(\bm r),v_\tau(\bm r),\{\Delta H_{R,ij}\})$: the
plane-wave potential, the potential of the kinetic-energy density and
the one-center Hamiltonian of each atom $R$. The operators
$\hat\rho_\alpha$ yield the conjugate densities $\bm\rho$. The
eigenstates of each trial Hamiltonian, with Fermi occupations for $N$
electrons, define a state for which we evaluate the Mermin free
energy~\cite{mermin65_pr137_A1441} $\mathcal A=E_{\mathrm{KS}}-TS$,
\begin{equation}
  \mathcal A(\vth)=\Omega(\vth)-\vth\cdot\bm\rho(\vth)
  +E_{\mathrm{pot}}[\bm\rho(\vth)] .
  \label{eq:A}
\end{equation}
Here $\Omega$ is the free energy of noninteracting electrons in
$\tilde H[\vth]$ at fixed electron number $N$, and $E_{\mathrm{pot}}$
contains the density-dependent contributions to the energy. Because
$\mathcal A$ is the free energy of an admissible ensemble, every trial
state provides an upper bound to $\mathcal A_{\min}$, the minimum of
the implemented functional over the states representable with the same
basis, PAW data sets, $k$-point sampling and symmetry. It is not a bound
to the exact free energy or to states outside this space. This remains true for approximate orbitals from an
iterative eigensolver, provided they are orthonormal with respect to
the PAW overlap operator. The gradient below assumes in addition that
the orbitals are exact eigenstates of $\tilde H[\vth]$, that the
occupations are smooth Fermi functions, so that $\Omega$ remains
differentiable at degeneracies, and that $E_{\mathrm{pot}}$ depends on
the state only through $\bm\rho$. The last condition holds for
semilocal functionals, meta-GGAs and DFT+$U$, but not for the
Fock-type terms of hybrid functionals (Supplementary Note~1, which
states the proposition and its assumptions). Since $\Omega$ is
concave, with $\partial\Omega/\partial\vth=\bm\rho$~\cite{gonze96_prb54_4383},
\begin{equation}
  \frac{\partial\mathcal A}{\partial\vth}=\chi\,(\vout-\vth),
  \qquad \vout=\frac{\partial E_{\mathrm{pot}}}{\partial\bm\rho},
  \label{eq:grad}
\end{equation}
where $\chi=\partial^2\Omega/\partial\vth^2$ is the symmetric,
negative-semidefinite static response. The residual $\bm r=\vout-\vth$
therefore satisfies
\begin{equation}
  \frac{\partial\mathcal A}{\partial\vth}\cdot\bm r
  =\bm r^{T}\chi\,\bm r\leq0 ,
  \label{eq:descent}
\end{equation}
so the residual is a non-ascent direction, and a strict descent
direction when it has a component outside the null space of $\chi$.
Besides the gauge mode below, null directions can arise, for example,
from linearly dependent projector functions.
Because $\chi$ is semidefinite, $\bm r^{T}\chi\,\bm r=0$ implies
$\chi\,\bm r=0$, so that the whole gradient then vanishes: a residual
in the null space of $\chi$, such as the gauge mode below, does not
change the state to first order. The
operator $\chi$ never has to be constructed. Near the minimum $\vth^*$,
$\bm r=-\varepsilon(\vth-\vth^*)$ with the dielectric operator
$\varepsilon=1-K_{\mathrm{Hxc}}\chi$ and the kernel
$K_{\mathrm{Hxc}}=\partial\vout/\partial\bm\rho$, so that the Newton
step is $\varepsilon^{-1}\bm r$. The inverse dielectric operator is thus
the natural preconditioner in Hamiltonian space, and the Kerker
filter~\cite{kerker81_prb23_3082} $G^2/(G^2+q_0^2)$, with $G=|\bm G|$
and the screening wave vector $q_0$, is its Thomas-Fermi model
(Supplementary Note~1).

\begin{figure}[!htbp]
\centering
\includegraphics[width=\linewidth]{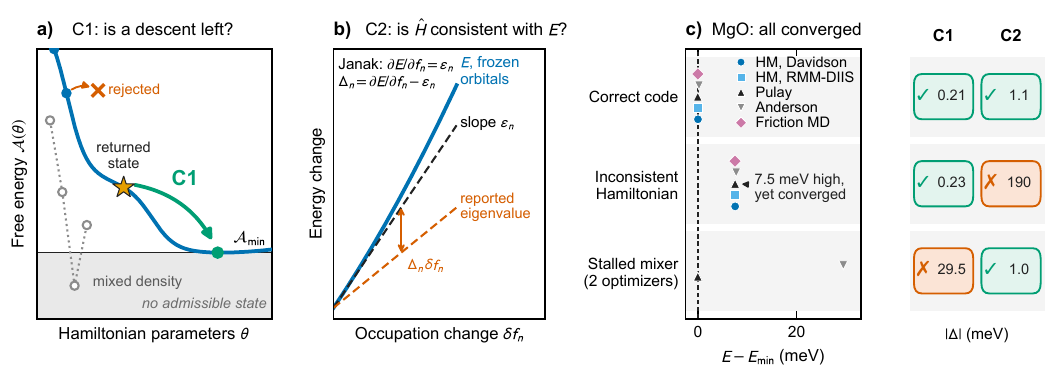}
\caption{\label{fig:certify}\textbf{The two tests.} a) C1, schematic: the
  free energy $\mathcal A(\vth)$ of the state induced by a trial
  Hamiltonian never falls below $\mathcal A_{\min}$ (shaded: no
  admissible state). HM accepts only steps that do not raise
  $\mathcal A$, whereas the energy reported for a mixed density (open
  circles) is not bounded. A returned state on a plateau or at an
  unstable stationary point can pass conventional criteria. The descent
  probe C1 restarts it variationally and exposes a remaining descent.
  b) C2, schematic: energy of frozen orbitals versus the occupation of
  one eigenstate. If the Hamiltonian is the derivative of the energy,
  the slope equals the eigenvalue $\epsilon_n$ [Eq.~(\ref{eq:janak})].
  An inconsistent Hamiltonian reports a different eigenvalue, and
  $\Delta_n\neq0$. c) MgO with SCAN0r at the $\Gamma$ point: final
  energies of the optimizers, all of which report convergence, relative
  to the minimum of the correct code, and the outcome of C1 (largest
  decrease, meV) and C2 (largest $|\Delta_n|$, meV), with thresholds
  $10\delta=0.27$~meV and 5~meV.}
\end{figure}

Hamiltonian minimization (HM) minimizes $\mathcal A(\vth)$ by Anderson
steps~\cite{anderson65_jacm12_547} along the preconditioned residual and
accepts a step only if $\mathcal A$ does not increase beyond the noise
of the eigensolver [Fig.~\ref{fig:certify}a]. Equation~(\ref{eq:descent})
concerns the residual itself. A preconditioned Anderson step is not
guaranteed to descend, which is why every step is tested on the energy. The energy reported
during density mixing, in contrast, is evaluated for a mixed density in
the spirit of the Harris-Foulkes functional~\cite{harris85_prb31_1770,
  foulkes89_prb39_12520}. Since the mixed density does not represent
the state induced by a trial Hamiltonian, the resulting energy is not
bounded by $\mathcal A_{\min}$. Related energy functions of the
potential have been used to choose the damping of potential
mixing~\cite{bendt82_prb26_3114,herbst22_jcp459_111127}, where steps
that raise the energy may still be accepted. In the PAW method a shift
$\tilde H\to\tilde H+c\hat O$ by the overlap operator leaves the state
unchanged. This null direction is shared between the plane-wave and the
one-center sectors and must be projected out of the residual.
Otherwise, the sectors drift apart and the iteration stalls above the
minimum (Supplementary Note~1). The implementation of HM and its
validation are summarized in Methods.

\reshead{Two tests}
Both tests apply to the state returned by any optimizer. They are
complementary rather than fully independent: C1 probes residual
variational descent, whereas C2 probes the consistency between
occupation derivatives of the implemented energy and the reported
eigenvalues. Both rest on the implemented energy, so an implementation
that is wrong but consistent can pass both.

\emph{C1: descent probe.} The returned state is restarted with a
variational optimizer, HM in VESPAW or direct minimization in other
codes, and we report
$\Delta\mathcal A_{\mathrm{C1}}=\mathcal A_{\mathrm{end}}-\mathcal A_{\mathrm{ret}}$.
The reference $\mathcal A_{\mathrm{ret}}$ is the free energy of the
state represented by the returned orbitals, evaluated in the first step
of the probe, not the energy that a density-mixing code reports for an
intermediate mixed density: only the former belongs to an admissible
state and obeys the upper bound. A decrease beyond the tolerance
demonstrates that the returned state is not the lowest state accessible
to the probe, and it provides a tighter upper bound on
$\mathcal A_{\min}$, provided that both energies refer to the same
implemented functional and numerical settings. If the decrease is continuous from the returned state, this is in addition evidence that the state is not a local minimum, within the directions and the resolution tested, either because it is not stationary or because it is an unstable stationary point. A trajectory with finite steps is not a complete stability analysis, and the evidence is strongest when the descent starts in the first steps of the probe. A
lower end point alone does not establish this: an optimizer can cross a
barrier and leave a metastable local minimum. We therefore also record
whether the energy falls steadily, without a rise exceeding the numerical tolerance, from the returned state until
the decrease exceeds the threshold. The converse is weaker: if the probe finds no
decrease, the returned state is a minimum only within the tolerance,
along the directions the probe explores, and for the Hamiltonian of the
code, whose errors C1 inherits.

\emph{C2: frozen-orbital Janak test.} C2 rests not on Hamiltonian
minimization but on Janak's theorem~\cite{janak78_prb18_7165}: if the
Hamiltonian is the functional derivative of the energy $E$, the
derivative of $E$ with respect to an eigenstate occupation, at fixed
orbitals, equals its eigenvalue,
\begin{equation}
  \Delta_n=\frac{\partial E}{\partial f_n}\Big|_{\psi}-\epsilon_n
  =\langle\psi_n|\hat H_E-\tilde H[\vth]|\psi_n\rangle ,
  \label{eq:janak}
\end{equation}
where $\hat H_E$ is the functional derivative of the implemented energy
and $\tilde H[\vth]$ the Hamiltonian whose eigenvalues are reported. The
derivative is taken at fixed occupations of all other states and
without the entropy term, which would add $-T\partial S/\partial f_n$.
The orbitals are frozen because relaxing them after the occupation
change would include the self-consistent response of the state. At
fixed orbitals, $\Delta_n$ isolates the derivative of the implemented
energy with respect to occupation. Because the orbitals are frozen, the
identity holds for each state individually, also within a degenerate
set, and for orbital-dependent terms such as Fock exchange and DFT+$U$,
provided $\epsilon_n$ is the generalized Kohn-Sham eigenvalue of the
Hamiltonian that the code uses. At a state that is not
self-consistent, $\hat H_E-\tilde H[\vth]$ also contains the residual,
so that C2 responds to incomplete relaxation as well. The tests are
therefore applied in sequence: C1 first, continued to its end point if
it lowers the free energy, and C2 at that end point, where a remaining
$\Delta_n$ indicates a Hamiltonian-energy inconsistency in the tested
state rather than incomplete relaxation. The numerical procedure and
the pass thresholds ($10\delta$ for C1, with $\delta$ the energy
tolerance per cell that the run claims to have met, and 5~meV for C2)
are given in Methods.

C1 cannot detect a Hamiltonian-energy inconsistency because the probe
shares the Hamiltonian, while C2 alone can miss incomplete relaxation in
degrees of freedom that do not affect the tested frontier states.
C2 is a differential test: it checks the Hamiltonian against the
implemented energy, not the energy against the intended functional.
Neither test can in general detect a functional that is implemented
incorrectly but consistently with its derivative, and passing both
does not establish that the state is the global minimum: on a nonconvex
landscape, even a genuine local minimum need not be the lowest one.

\reshead{Controlled failures}
Two errors found during the development of VESPAW were reintroduced to
test C1 and C2 on MgO with the local hybrid SCAN0r at the $\Gamma$
point (Fig.~\ref{fig:certify} and Table~\ref{tab:certify}), with the
default tolerance $\delta=10^{-6}$~Hartree (0.027~meV). In the first,
the double-counting term of the onsite hybrid subtracted the
kinetic-energy-density potential of the frozen core. The energy
functional is unchanged, but the Hamiltonian is not its derivative. All
five optimizers converge to the same wrong state. The descent probe
passes: because it shares the Hamiltonian, C1 cannot detect this
inconsistency. The Janak test fails by $-92$ and $-190$~meV for the
lowest unoccupied and highest occupied states, compared with about
1~meV for the correct code.

In the second error, the density residual of the Pulay and Anderson
mixers included Fourier components outside the cutoff sphere, which the
Kerker filter never mixes. The residual stayed at a floor, the mixers
stopped on the change of the energy, and Anderson mixing ended 29~meV
above the minimum with an unrelaxed one-center density matrix. Its
frontier orbitals are converged, so C2 passes, but C1 lowers the free
energy by 29.45~meV in two accepted steps. With both errors removed, all
optimizers end within 0.21~meV of the minimum and pass both tests. The
two controlled failures are of opposite type, and each is caught by
exactly one of the tests.

\begin{table}[!htbp]
\caption{\label{tab:certify}\textbf{Tests in two codes, grouped by
  outcome.} C1: $\Delta\mathcal A_{\mathrm{C1}}$, the change of the free
  energy of the returned state when it is restarted with a variational
  optimizer (HM in VESPAW, \texttt{ALGO=All} in VASP). C2: $\Delta_n$ of
  Eq.~(\ref{eq:janak}) for the lowest unoccupied (L) and highest
  occupied (H) state. All calculations report convergence. Bold: failed
  (C1: decrease $>10\delta$, with $\delta$ the requested energy
  tolerance per cell, and C2: $|\Delta_n|>5$~meV). Energies in meV.}
\centering
\begin{tabular}{lrr}
\toprule
System & C1 & C2 (L/H) \\
\midrule
\multicolumn{3}{l}{\emph{Correct: both tests pass}}\\
MgO, VESPAW, 5 optimizers$^{a}$       & $\geq-0.21$ & $1.1/0.3$ \\
MgO, VASP, 5 functionals$^{b}$         & ---         & $|\Delta|\leq3.5$ \\
NiO, VASP, \texttt{LMAXMIX=4}          & $-0.00004$  & $-0.9/-1.7$ \\
LiMn$_2$O$_4$, VASP                    & $-0.12$    & $1.6/0.0$ \\
15 oxides, VASP, database$^{c}$        & $\geq-0.71$ & --- \\
\midrule
\multicolumn{3}{l}{\emph{Hamiltonian inconsistency: C2 fails}}\\
MgO, VESPAW, 5 optimizers$^{a}$       & $\geq-0.23$ & $\mathbf{-92/-190}$ \\
\midrule
\multicolumn{3}{l}{\emph{Descent remains: C1 fails}}\\
MgO, VESPAW, Anderson                  & $\mathbf{-29.45}$ & $0.5/-0.8$ \\
NiO, VASP, \texttt{LMAXMIX=2}          & $\mathbf{-0.030}$ & $-0.9/-0.8$ \\
UO$_2$, VASP, $\Gamma^{d}$           & $\mathbf{-1460}$ & $-0.7/1.0^{e}$ \\
UO$_2$, VASP, $2\times2\times2$ $k^{d}$ & $\mathbf{-212}$ & --- \\
Ce$_2$O$_3$, VASP, $\Gamma^{d}$      & $\mathbf{-17.57}$ & $0.3/\mathbf{7.3}^{e}$ \\
FeO, VASP, database$^{c}$              & $\mathbf{-899}$ & --- \\
CoO, VASP, database$^{c}$              & $\mathbf{-789}$ & --- \\
Fe$_3$O$_4$, VASP, database$^{c}$      & $\mathbf{-239}$ & --- \\
4 oxides, VASP, database$^{c,f}$       & $\mathbf{\leq-5.5}$ & --- \\
\bottomrule
\end{tabular}

\raggedright\footnotesize $^{a}$Largest magnitude over HM (Davidson,
RMM-DIIS), Pulay, Anderson and friction dynamics. $\delta=0.027$~meV.
$^{b}$PBE, r$^2$SCAN, SCAN, HSE06, PBE0. Largest $|\Delta_n|$ for SCAN.
$^{c}$Materials Project structures and static protocol, Gaussian smearing. $\delta=N_{\mathrm{at}}\times5\times10^{-5}$~eV.
$^{d}U$ ramped from PBE over $U/2$ to $U$, \texttt{LMAXMIX=6}, $\delta=10^{-7}$~eV. For UO$_2$ at $\Gamma$, C1 is $-1436$~meV with
$\delta=10^{-4}$ and $10^{-6}$~eV.
$^{e}$At the returned state. At the end point of C1, $-1.0/1.5$
(UO$_2$) and $-0.5/1.0$ (Ce$_2$O$_3$).
$^{f}$LiFePO$_4$ $-46$, V$_2$O$_3$ $-14$, Co$_3$O$_4$ $-9.8$, FeTiO$_3$
$-5.5$.
NiO: 64 atoms, $\delta=10^{-7}$~eV. LiMn$_2$O$_4$: 56 atoms,
$\delta=10^{-4}$~eV.
\end{table}

\reshead{False convergence in another code}
Both tests apply to VASP~6.4.2~\cite{kresse96_prb54_11169,
  kresse99_prb59_1758} without changes to the code
(Table~\ref{tab:certify}, procedure in Methods). For MgO, the Janak test
passes for five functionals from PBE to the Fock hybrids, showing no
Hamiltonian-energy inconsistency in the tested frontier states. For
antiferromagnetic NiO, Pulay mixing with the default
\texttt{LMAXMIX=2} passes only the angular components up to $L=2$ of
the one-center densities to the mixer and reports convergence with
$\delta=10^{-7}$~eV. C1 nevertheless lowers the free energy of the
returned orbitals by 30~$\mu$eV, $300\delta$, and with
\texttt{LMAXMIX=4} it passes. The decrease is too small to matter
physically, but it shows that the probe resolves, in another code, the
signature of the residual floor in VESPAW: a consistent state from
which a descent remains, not detected by the convergence criterion or
the Janak test but detected by the descent probe.

\begin{figure}[!htbp]
\centering
\includegraphics[width=0.6\linewidth]{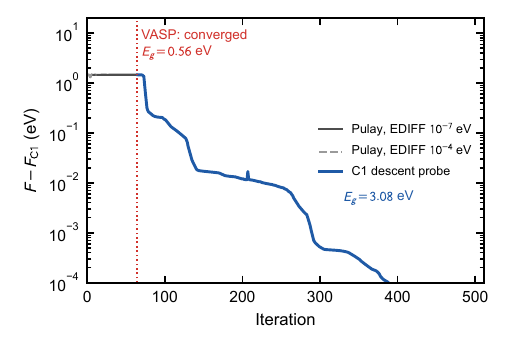}
\caption{\label{fig:uo2}\textbf{False convergence of UO$_2$ with PBE+$U$
  in VASP~6.4.2.} Free energy relative to the end of the descent probe
  C1. Pulay mixing (grey, \texttt{LMAXMIX=6}, after a $U$ ramp) reports
  convergence (dotted line) with a band gap of 0.56~eV. C1, started from
  the same orbitals (blue, \texttt{ALGO=All}), lowers the free energy by
  1.46~eV and opens the gap to 3.08~eV.}
\end{figure}

A much larger C1 failure occurs for UO$_2$ (Fig.~\ref{fig:uo2}).
PBE+$U$ calculations of $f$-electron oxides are known to end in
metastable orbital states~\cite{dorado09_prb79_235125}, and $U$ is
commonly ramped up to steer the iteration. In an antiferromagnetic cell
with 12 atoms, ramped from PBE to $U-J=3.96$~eV with
\texttt{LMAXMIX=6}, Pulay mixing reports convergence with tolerances
from $10^{-4}$ to $10^{-7}$~eV, always at the same state, which is
self-consistent according to the mixer. The descent probe, started from
the same orbitals, lowers the free energy by 1.44 to 1.46~eV and ends in
an insulating state with a gap of 3.08~eV instead of 0.56~eV. The first
step of the descent changes the energy by only 0.5~$\mu$eV, and the
decrease then grows by six orders of magnitude within 11 steps. The
returned state therefore suggests a nearly stationary, variationally
unstable solution: a mixing scheme can come to rest there, whereas a
variational descent departs. Damped electronic dynamics started from a
recomputed returned state of the same kind departs in the same way for
three time steps, whereas in NiO it decays (Supplementary Note~9). Residual and energy criteria cannot
distinguish such a state from a minimum, but C1 can. The Janak test
passes within 1~meV at the returned state and within 1.5~meV at the end
point of C1: the returned state passes the consistency test, but is not
a minimum. The error is not an artifact of the $\Gamma$ point: with a
$2\times2\times2$ $k$-point mesh the same protocol reports convergence
0.21~eV above the end point of C1. The probe does not find the ground
state. The probes from the three $\Gamma$-point converged states end in
two different minima, 24~meV apart, and a probe from random orbitals
reaches a state at least 14~meV lower still.

The error need not be large to be systematic. For ferromagnetic
Ce$_2$O$_3$ with one $4f$ electron per Ce and 20 atoms, the same
protocol converges in 42 iterations with $\delta=10^{-7}$~eV. The
descent probe lowers the free energy by 17.6~meV, more than
$10^{5}\delta$, while the band gap stays at 2.5~eV: the occupied $f$
orbitals rotate slowly in a flat part of the landscape that the mixer
does not resolve. At the returned state the Janak test fails as well,
with 7~meV for the highest occupied state, because the state is not
fully relaxed and C2 therefore contains a residual contribution. At
the end point of the probe it passes within 1~meV. This illustrates why
C2 is interpreted only after C1.

\reshead{Targeted survey under a database protocol}
Can such failures occur under a production database protocol? We took
22 magnetic oxides from the Materials
Project~\cite{jain13_aplmater1_011002}, among them rocksalt, corundum,
rutile, spinel and perovskite oxides and the cathode materials
LiCoO$_2$, LiNiO$_2$, LiMn$_2$O$_4$ and LiFePO$_4$, and recomputed them
in their database structures with its static
protocol~\cite{ong13_cms68_314,jain11_prb84_045115} (Methods). All 22
report convergence. Fifteen pass C1 within $2.1\delta$, and seven fail
(Table~\ref{tab:survey}).

The failures show predominantly two behaviors. For rocksalt CoO and FeO, the
protocol returns metallic states with gaps below 5~meV. C1 lowers their
free energies by 0.79 and 0.90~eV at unchanged total moment and ends in
insulating states with gaps of 1.40 and 1.38~eV, respectively. For
Fe$_3$O$_4$, it lowers the free energy by 0.24~eV and opens the gap
from 0.03 to 0.49~eV, and for LiFePO$_4$ it lowers it by 46~meV. In
these four cases the iteration oscillates by several meV per step and
stops when a single energy change happens to fall below the tolerance.
In CoO, for example, the last three changes are $+3.4$, $-0.70$ and
$+0.42$~meV before a change of $+0.06$~meV ends the run at
$\delta=0.1$~meV. The returned state of CoO is also close to an
unstable point: over nine successive iterations of C1, the decrease per
iteration grows by a factor of 1.9 to 2.5 per step, from 0.14 to
79~meV. Damped electronic dynamics with three time steps leaves the
returned state in the same way and reaches the same end point. In V$_2$O$_3$, Co$_3$O$_4$ and FeTiO$_3$, the
iteration is still descending steadily, by 0.3 to 4~meV per step, when
a single small energy change satisfies the stopping criterion. C1 lowers
the free energy by 14, 9.8 and 5.5~meV, 29, 14 and $11\delta$, and
opens the gap of V$_2$O$_3$ from 0.13 to 0.64~eV.

With experimental structures instead, the same protocol lets these
seven pass and LiCoO$_2$ fail by 0.37~eV (Supplementary Note~7), so the
outcome depends on the starting structure and not only on the material.
The 22 systems were chosen as difficult magnetic oxides. The targeted
survey therefore shows that such failures occur under a production
protocol, but not how often they occur.

\begin{table}[!htbp]
\caption{\label{tab:survey}\textbf{Targeted survey with the Materials
  Project static protocol} in VASP~6.4.2, database structures. $N$:
  atoms in the cell. $\delta$: energy tolerance per cell (meV). it.:
  iterations to the reported convergence. $\Delta\mathcal A_{\mathrm{C1}}$:
  change of the free energy of the returned orbitals under C1 (meV).
  $M$: total moment ($\mu_B$) before and after C1. Bold: C1 failed
  ($-\Delta\mathcal A_{\mathrm{C1}}>10\delta$). All runs report
  convergence.}
\centering\small
\begin{tabular}{llrrrrrr}
\toprule
System & Entry & $N$ & $\delta$ & it. & $\Delta\mathcal A_{\mathrm{C1}}$ & $-\Delta\mathcal A_{\mathrm{C1}}/\delta$ & $M$\\
\midrule
NiO & mp-19009 & 2 & $0.1$ & 14 & $-0.002$ & $0.0$ & $2.00\to2.00$ \\
MnO & mp-1232659 & 2 & $0.1$ & 11 & $-0.014$ & $0.1$ & $5.00\to5.00$ \\
FeO & mp-1274279 & 2 & $0.1$ & 17 & $\mathbf{-898.9}$ & $\mathbf{8989.4}$ & $4.01\to4.00$ \\
CoO & mp-19079 & 2 & $0.1$ & 21 & $\mathbf{-789.3}$ & $\mathbf{7892.6}$ & $3.00\to3.00$ \\
CuO & mp-704645 & 4 & $0.2$ & 13 & $-0.055$ & $0.3$ & $0.03\to0.00$ \\
LiCoO$_{2}$ & mp-22526 & 4 & $0.2$ & 14 & $-0.226$ & $1.1$ & $0.00\to0.00$ \\
LiNiO$_{2}$ & mp-2348641 & 12 & $0.6$ & 17 & $-0.701$ & $1.2$ & $3.00\to3.00$ \\
SrFeO$_{3}$ & mp-510624 & 5 & $0.25$ & 15 & $-0.514$ & $2.1$ & $4.03\to4.03$ \\
SrCoO$_{3}$ & mp-1283849 & 10 & $0.5$ & 17 & $-0.351$ & $0.7$ & $6.19\to6.19$ \\
MnO$_{2}$ & mp-510408 & 6 & $0.3$ & 13 & $-0.109$ & $0.4$ & $6.00\to6.00$ \\
CrO$_{2}$ & mp-19177 & 6 & $0.3$ & 15 & $-0.216$ & $0.7$ & $4.00\to4.00$ \\
VO$_{2}$ & mp-19094 & 6 & $0.3$ & 14 & $-0.041$ & $0.1$ & $2.00\to2.00$ \\
Cr$_{2}$O$_{3}$ & mp-19399 & 10 & $0.5$ & 13 & $-0.078$ & $0.2$ & $12.00\to12.00$ \\
Fe$_{2}$O$_{3}$ & mp-19770 & 10 & $0.5$ & 14 & $-0.123$ & $0.2$ & $20.00\to20.00$ \\
V$_{2}$O$_{3}$ & mp-18937 & 10 & $0.5$ & 16 & $\mathbf{-14.353}$ & $\mathbf{28.7}$ & $8.00\to8.00$ \\
FeTiO$_{3}$ & mp-1277107 & 10 & $0.5$ & 27 & $\mathbf{-5.534}$ & $\mathbf{11.1}$ & $8.00\to8.00$ \\
Fe$_{3}$O$_{4}$ & mp-19306 & 14 & $0.7$ & 16 & $\mathbf{-238.6}$ & $\mathbf{340.9}$ & $28.01\to28.00$ \\
Co$_{3}$O$_{4}$ & mp-1271793 & 14 & $0.7$ & 18 & $\mathbf{-9.767}$ & $\mathbf{13.9}$ & $6.00\to6.00$ \\
MnCr$_{2}$O$_{4}$ & mp-28226 & 14 & $0.7$ & 13 & $-0.089$ & $0.1$ & $22.00\to22.00$ \\
ZnFe$_{2}$O$_{4}$ & mp-19313 & 14 & $0.7$ & 14 & $-0.024$ & $0.0$ & $20.00\to20.00$ \\
LiMn$_{2}$O$_{4}$ & mp-1272804 & 14 & $0.7$ & 15 & $-0.462$ & $0.7$ & $14.00\to14.00$ \\
LiFePO$_{4}$ & mp-19017 & 28 & $1.4$ & 27 & $\mathbf{-46.309}$ & $\mathbf{33.1}$ & $16.01\to16.00$ \\
\bottomrule
\end{tabular}
\end{table}

\reshead{Prevalence in a census of database structures}
To measure how often a calculation that reports convergence under this
protocol retains a variational descent, we computed all 366 experimentally observed
magnetic oxides of V, Cr, Mn, Fe, Co, Ni and Cu in the Materials
Project with an energy above the hull of at most 25~meV/atom and at
most 20 sites in the database cell, excluding the 22 compounds above
(Methods, Supplementary Table~14). The population was fixed before any
calculation was run. Because the census enumerates the whole
population, its rates for this population have no sampling error. This
does not make them transferable: they characterize the modified
protocol applied to these structures, not the energies stored in the
database, and not other classes of materials, protocols or codes.

The 23 entries that contain W are treated separately, a decision taken
after the census had revealed their anomaly and before any control was
run. With the potential \texttt{W\_pv}, 15 of the 21 that report
convergence fail C1, but these runs are anomalous: most end with a
charge-density residual far above that of any other entry, and in ten
the free energy of the returned orbitals lies 0.3 to 28~eV above the
reported energy. We repeated the 23 entries with \texttt{W\_sv} and
otherwise identical settings, having fixed in advance that the rate
including W would be the main result only if at least 8 of the 15
failures persisted. The anomalies disappear, all 23 report convergence,
and only 3 of the 15 failures persist, while three other entries fail
(Supplementary Note~8, Supplementary Table~13). The comparison
implicates the \texttt{W\_pv} potential as used here, but it does not
establish the cause, because the two data sets also differ in their
valence (12 and 14 electrons) and release. We give the rates below for
the 343 entries without W and treat W as a sensitivity analysis:
including the W entries, the rate at $10\delta$ is 20.0\% with
\texttt{W\_pv} and 17.4\% with \texttt{W\_sv}.

\begin{figure}[!htbp]
\centering
\includegraphics[width=\linewidth]{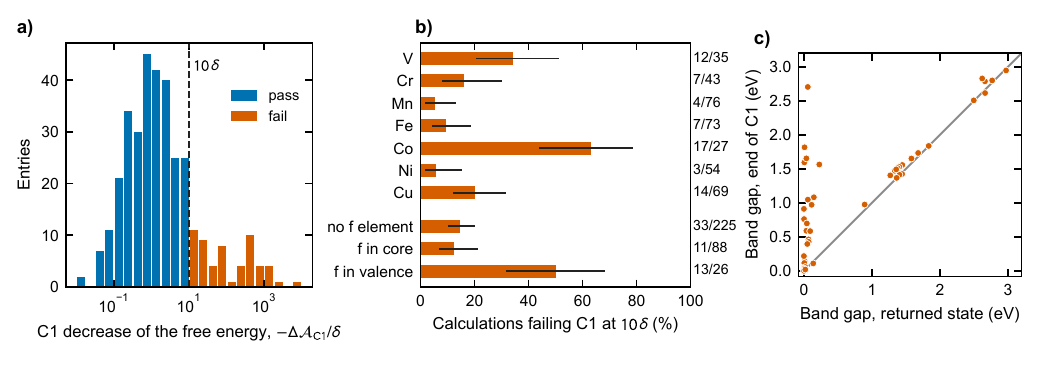}
\caption{\label{fig:prevalence}\textbf{Prevalence of false convergence
  in a census of Materials Project magnetic oxide structures.} The 339
  calculations without W that report convergence under the modified
  static protocol. a)~Distribution of the decrease of the free energy in C1 in
  units of the requested tolerance $\delta$, with the threshold
  $10\delta$. Values below $10^{-2}$ are placed in the lowest bin.
  b)~Fraction of the converged calculations that fail C1 at $10\delta$, by transition-metal
  element and by the treatment of $f$ electrons, with Wilson 95\%
  intervals and counts. Elements overlap in compounds containing more
  than one. c)~Band gap of the returned state and of the end point of
  C1 for the 56 failures in which C1 converges. The 57th, whose C1 does
  not converge within 1000 iterations, is omitted. The line marks an
  unchanged gap.}
\end{figure}

Of the 343 entries, four do not report convergence within 200
iterations, a visible failure that needs no test. Of the 339 that do,
86 (25.4\%) fail C1 at $5\delta$, 57 (16.8\%) at $10\delta$, 37
(10.9\%) at $30\delta$, 25 (7.4\%) at $100\delta$ and 23 (6.8\%) at
$300\delta$ (Fig.~\ref{fig:prevalence}a). In 49 of the 57 failures at
$10\delta$ (14.5\% of the 339), the energy falls steadily from the returned state, with no rise larger than $\delta$, until the decrease
exceeds $10\delta$. In the other eight it first rises, so that the end
point shows a lower state but not by itself an instability of the
returned one. Along the whole trajectory, 19 of the 57 show a rise
larger than $\delta$. For 23 entries
(6.8\%) the reported free energy lies more than 100~meV above the end
point of C1, and for nine (2.7\%) more than 0.5~eV. Per atom, the error
exceeds 25~meV, the width of the stability window used to select these
entries, for 12 (3.5\%).

The failures are not only errors of the energy. Among the 56 failures
in which C1 converges, 18 open the band gap by more than 0.2~eV and 12
by more than 0.5~eV. Ten change from a metal (gap below 0.1~eV) to an
insulator (gap above 0.5~eV, Fig.~\ref{fig:prevalence}c), and 19 change
the total moment by more than 0.1~$\mu_{\mathrm B}$. Thirty show at
least one of these changes. Of the 282 entries that pass, none opens
the gap by more than 0.2~eV, and two change the moment. In the largest
failure, GdCrTeO$_6$, the reported energy lies 5.8~eV above the end
point of C1, the moment changes from 3.25 to 8.00~$\mu_{\mathrm B}$ and
the gap from 0.06 to 2.71~eV.

The rate depends strongly on the chemistry (Fig.~\ref{fig:prevalence}b).
It is 63\% of the compounds containing Co (17 of 27) and 34\% of those
containing V, but 5--6\% of those containing Mn or Ni. It is 50\% (13 of
26) for compounds with $f$ electrons in the valence (Ce, Eu, Gd, U and
Np, without a Hubbard correction on the $f$ shell), and 14.1\% for the
others, including 12.5\% for those whose $f$ electrons are in the
frozen core. It rises from 11.8\% for computed cells of up to 10 atoms
to 19.0\% for cells of 16 to 20 atoms. These rates hold for this
protocol and initialization and do not by themselves show that the
chemistry causes the failures. All runs start from ferromagnetic moments,
whatever the ordering in the database, so that the census tests this
initialization and not the database magnetic order: none of the 42
entries labelled antiferromagnetic ends with a net moment below
0.1~$\mu_{\mathrm B}$, and for 12 of them the chemically primitive cell
used here is smaller than the database cell and cannot hold that
order (a control from the database order follows below). The rate is 14.4\% (39 of 271) for entries
labelled ferromagnetic, 16.7\% (7 of 42) for antiferromagnetic and
42.3\% (11 of 26) for ferrimagnetic. These groups overlap and are not
independent. Neither the energy history nor the residual reported by
the code identifies the failures: two or more sign changes among the
last six energy changes occur in 37\% of the failures and 34\% of the
passes, and the median charge-density residual is 0.024 and 0.020,
respectively.

Four controls test the protocol (Methods, Supplementary
Note~9). First, the starting state: for 25 failures and 25
passes drawn at random, the run was repeated with the tetrahedron method
of the original input set and probed with C1 with Gaussian smearing as
before. The tetrahedron run minimizes a different free energy, so this
control does not test whether it reached a minimum of its own functional. It tests whether the failures depend on the state from
which the probe starts. Of the 25 failures, 22 still fail at
$10\delta$, and 2 of the 25 passes fail, so the failures are not
specific to a start from a run with Gaussian smearing. Second, the
magnetic start: the 42 entries labelled antiferromagnetic were repeated
from the moments and cell of the database. Of these, 41 report
convergence and all 41 end with a net moment below
0.1~$\mu_{\mathrm B}$. The end point of C1 lies below that of the
ferromagnetic start in 40 of the 42 entries, by a median of
3.6~meV/atom. Four of the 41 fail C1 at $10\delta$ (9.8\%), compared
with 7 of 42 from the ferromagnetic start: two failures become passes,
and no pass becomes a failure. Third, the optimizer: for 20 failures
drawn at random and for CoO and FeO, Pulay mixing started from the end
point of C1 reports convergence within two iterations, within 2.6~meV
of the free energy of C1 and with a gap changed by at most 0.07~eV, so
the end points are not an artifact of direct minimization. Fourth, a plain restart. All 57 failures and 25 randomly drawn passes
were computed again and then restarted from their orbitals with
unchanged settings. The repeated runs give the same result for 56 of
the 57 failures and for all 25 passes. The one exception,
Na$_3$Co$_2$SbO$_6$, now gives $7.5\delta$ and passes, so the repeated
set contains 56 failures and 26 passes. The restart ends more than
$10\delta$ below the reported free energy for only 15 of the 56
failures (27\%) and for none of the passes. We had fixed 80\% in
advance as the condition for a restart to replace C1. In 33 of the 41
missed failures the restart ends within $3\delta$ of the reported free
energy, even where C1 lowers the energy by more than $800\delta$, and
two restarts do not converge. Restarting with the same algorithm thus
brings the calculation back to almost the same state. The
first control also shows the limit of a pass. Started from the run
with the tetrahedron method, C1 for CoO ends in a state with a gap of
0.02~eV and passes, although its free energy lies 0.60~eV above the
insulating end point reached from the run with Gaussian smearing. Both end points are evaluated with the same functional and settings of
C1.

\reshead{Cost}
C2 requires five frozen-orbital energy evaluations. In the census, C1
takes a median of 14 iterations for a state that passes and 68 for one
that fails, compared with 15 for the SCF run. Its median wall time is
0.84 of that of the SCF run it tests (interquartile range 0.67--1.13):
0.80 for states that pass and 2.3 for those that fail, up to 17 times
the SCF run. Applying the full probe therefore about doubles the
median cost of a static calculation in this census, and it costs most where it is most needed. Two modes of use
differ in cost. As an early warning, the probe stops as soon as the
decrease exceeds $10\delta$, which takes a median of 7 iterations for
the failures, and a state that passes costs the 0.80 given above. As a
full correction, the probe continues to the lower end point, at the
median cost of 2.3 static runs given above for the failures. Relative
to a workflow that relaxes the structure before the static run, the
additional cost is smaller. A plain restart is cheaper, a median of 0.22 of
the run against 1.39 for C1 in the same set, but it misses 41 of the 56
failures.

\section*{Discussion}
The examples establish a hierarchy. A small change of the energy does
not imply convergence, a small residual does not imply a correct
solution, and agreement between independent optimizers does not imply
either when they share the Hamiltonian. What the variational
formulation adds is a pair of falsifiable statements about a returned
state: that a variational restart finds no descent beyond the tolerance
(C1), and that, at the end point of C1, the tested frontier states show
no inconsistency between the reported Hamiltonian and energy (C2). A
failure of C1 shows that a lower state is accessible from the
returned state and, if the descent is continuous, it gives evidence that the returned state is not a local minimum within the directions and the resolution tested. A failure of C2 at the end point of C1
indicates a Hamiltonian-energy inconsistency in the tested state. A
pass holds relative to the probe, its tolerance and the Hamiltonian
space of the code. The first test follows from the upper-bound property
of the free energy, with Eqs.~(\ref{eq:grad}) and (\ref{eq:descent})
providing the descent direction of HM, and the second from
Janak's theorem at frozen orbitals. Together they separate two distinct
ways in which a calculation can falsely appear converged: a remaining
descent and an inconsistent Hamiltonian, which no single conventional
criterion distinguishes. The mechanisms behind the failures are
classified in Table~\ref{tab:mech}.

\begin{table}[!htbp]
\caption{\label{tab:mech}\textbf{Mechanisms of false convergence in this
  work.} The interpretation of UO$_2$ and CoO as close to unstable
  points rests on the trajectories of C1 and of damped electronic
  dynamics, in which the residual grows roughly geometrically
  over the first iterations from the returned state, while that of NiO decays. A quantitative criterion fixed in
  advance was not met, so the interpretation is qualitative
  (Supplementary Note~9). The mechanisms are not exclusive, and CoO shows
  two.}
\centering\small
\begin{tabular}{>{\raggedright\arraybackslash}p{0.26\linewidth}>{\raggedright\arraybackslash}p{0.18\linewidth}>{\raggedright\arraybackslash}p{0.29\linewidth}>{\raggedright\arraybackslash}p{0.17\linewidth}}
\toprule
Mechanism & Examples & Signature & Underlying issue \\
\midrule
Accidental small energy change & FeO, Fe$_3$O$_4$, LiFePO$_4$, CoO &
Energy oscillates by meV, one small change ends the run & Stopping
criterion \\
Incomplete relaxation & V$_2$O$_3$, Co$_3$O$_4$, FeTiO$_3$, Ce$_2$O$_3$ &
Energy still descends, or orbitals rotate slowly, when the run stops &
Premature termination \\
Nearly stationary unstable state & UO$_2$, CoO & C1 begins with a
minute change that grows geometrically & Variational stability \\
Degrees of freedom missing from the mixer & MgO (Anderson), NiO
(\texttt{LMAXMIX=2}) & Residual floor, C1 fails and C2 passes & Mixer \\
Hamiltonian-energy inconsistency & MgO (double counting) & Optimizers
agree, C1 passes and C2 fails & Implementation \\
\bottomrule
\end{tabular}
\end{table}

For databases and the models trained on them, the relevant failure is
not the energy alone. In CoO, FeO, Fe$_3$O$_4$ and V$_2$O$_3$ the
falsely converged state differs qualitatively from the end point of the
probe: a metal instead of an insulator, or a gap that changes by a
factor of five or more. Such entries can propagate the artifact into
models trained on them. In the census, more than half of the failures with a
converged probe change the gap or the moment, and the failures
concentrate in compounds of Co and V and in those with $f$ electrons in
the valence. Because the tests require only a restart from the stored
orbitals, they can also be applied retrospectively to existing entries
whose wave functions or inputs are available. The census rates refer to
the modified protocol described in Methods, with Gaussian smearing and a
ferromagnetic start from scratch, and they are not error rates of the
stored database entries, and they count only the calculations that
report convergence. Started from runs with the tetrahedron method, 22
of 25 sampled failures persist under the same probe. With the antiferromagnetic order of the database, the
rate of those entries falls from 7 of 42 to 4 of 41.

Testing whether a self-consistent solution is a minimum has a long
history in quantum chemistry, where the orbital Hessian of Hartree-Fock
and Kohn-Sham solutions is examined for negative
eigenvalues~\cite{seeger77_jcp66_3045,bauernschmitt96_jcp104_9047},
and direct minimization reaches minima of the energy rather than fixed
points of the mixing~\cite{payne92_rmp64_1045,marzari97_prl79_1337,
  freysoldt09_prb79_241103,cances00_ijqc79_82,woods19_jpcm31_453001}.
C1 needs no Hessian: it uses the variational optimizer that the code
already has and only the energy of the returned state, so it applies to
plane-wave PAW calculations with many states and to the output of any
mixing scheme. Unlike a Hessian analysis, it does not classify the
stationary point. A decrease beyond the tolerance shows a lower
accessible state, and only the shape of the trajectory indicates an
instability.

For automated workflows, this suggests five practical rules.
(i)~After a run that reports convergence, restart from its orbitals
with a variational optimizer (in VASP, \texttt{ALGO=All}), compare
with the free energy of the returned orbitals, evaluated in the first
step, and flag the run if the decrease exceeds $10\delta$. For an early warning, stop there, and for a full correction continue to the end
point.
(ii)~For new or modified functionals and code versions, apply the
frozen-orbital Janak test at the end point of C1 and flag $|\Delta_n|>5$~meV.
(iii)~In the correction mode, store the end point of C1, not the
returned state, when C1 fails, after checking that C1 converged with
the settings of the production run.
(iv)~Do not rely on a tighter tolerance or on agreement between mixing
schemes: the UO$_2$ calculation reports the same false state from
$10^{-4}$ to $10^{-7}$~eV. A plain restart with the same
settings does not replace C1 either: in the census it misses 41 of 56
failures. (v)~Sample competing magnetic and orbital
orders separately, since the tests do not address the global minimum.

Three limits are essential. First, passing C1 and C2 does not prove
that the state is the global minimum. Competing magnetic or charge
orders must still be sampled, and in UO$_2$ and LiCoO$_2$ the probe
ends in different minima from different starting points. Started from
a run with the tetrahedron method, C1 for CoO passes a nearly gapless
state 0.60~eV above the insulating state of the same functional: a pass
excludes a remaining descent, not a false state. Second, an
error that changes the functional and its derivative consistently can
pass both tests. In our development, a missing factor 1/2 in the
one-center kinetic-energy density of the meta-GGA passed a relaxed
Janak test within 5~meV and was found only by comparison with another
code (Supplementary Note~6). Third, C2 probes only the frontier states,
and C1 only the Hamiltonian space of the code, including its symmetry:
a state that is stable under symmetric variations can be unstable when
the symmetry is broken.

In summary, treating the Kohn-Sham Hamiltonian as the variable of an energy minimization, in which the free energy of the induced state is
an upper bound and the residual weighted by the response function is
its gradient, supplements conventional convergence criteria with
falsifiable tests of variational descent and of Hamiltonian-energy
consistency. The resulting descent probe, together with a
frozen-orbital Janak test, exposes false convergence that conventional
criteria and agreement between optimizers can miss. In VASP, it finds a
UO$_2$ calculation converged 1.46~eV above a lower state and, under a
database protocol, CoO and FeO returned as metals 0.8--0.9~eV above
insulating states, and it finds a decrease beyond the threshold in
16.8\% of the converged calculations in a census of tungsten-free
Materials Project structures recomputed with a modified static
protocol. The tests require no
change of the code to which they are applied.

\begin{methods}
\runin{Hamiltonian minimization.}
We implemented HM in the PAW code VESPAW~\cite{sotoudeh26_vespaw_zenodo},
derived from CP-PAW~\cite{bloechl94_prb50_17953}. Steps are proposed by
an Anderson extrapolation~\cite{anderson65_jacm12_547} from the
accepted points $(\vth_k,\bm r_k)$, applied to the preconditioned
residual with a damping factor $\beta$,
$\vth'=\vth_1+\beta P\bm r_1-\sum_{k\ge2}\gamma_k[(\vth_1-\vth_k)+\beta
P(\bm r_1-\bm r_k)]$, where the coefficients $\gamma_k$ minimize the
norm of the extrapolated residual and $P$ is the Kerker preconditioner.
If the extrapolated step exceeds the plain damped step $\beta P\bm r_1$
by more than a factor of ten, the plain step is used. For each trial
$\vth$, the eigenstates of $\tilde H[\vth]$ are determined with the
Davidson, RMM-DIIS or conjugate-gradient eigensolver, the Fermi
occupations are set for fixed $N$, $\mathcal A(\vth)$ is evaluated for
this state, and the gauge mode $\tilde H\to\tilde H+c\hat O$ is
projected out of the residual. A proposal is accepted if
$\mathcal A(\vth')\le\mathcal A_{\mathrm{best}}+\epsilon$, with a slack
$\epsilon=10^{-7}$~Ha that absorbs the noise of the iterative
eigensolver. Otherwise the step is halved towards the best point, and
the best point is evaluated again after five consecutive rejections.
The damping $\beta\in[0.02,0.5]$ is halved once per sequence of
rejections, at the second consecutive rejection, when the Anderson
history is also discarded, and increased by a factor 1.2 after an
accepted step and 1.5 after three consecutive accepted steps. The
iteration is converged when the mean-square residual of the grid
potential is below its threshold and the largest residual of the
one-center Hamiltonian below its square root, or after two consecutive
accepted full-length steps whose energy gain is below the energy
tolerance. In cells with vacuum, $\vth$ is set equal to $\vout$ where
the density is below $10^{-6}\,a_B^{-3}$, and $\vth$ and the history are
shifted along the gauge mode when its offset from $\vout$ exceeds
0.5~Ha. For the local hybrid functionals, the onsite Fock-type
correction is applied as an operator constructed from the current
state. The eigensolvers return orthonormal eigenstates: the Davidson
correction vectors are orthogonalized twice to all current bands and
the extended basis is orthogonalized canonically with an eigenvalue
cutoff of $10^{-10}$, and RMM-DIIS is preceded by eight Davidson
iterations. With space-group symmetry, the output density and the
one-center density matrices are symmetrized, so that every residual
and every Anderson combination keeps $\vth$ in the symmetric subspace.
In parallel runs, all scalar products of the energy test and the
Anderson fit are summed over the MPI tasks, so that every task takes the
same decisions. The derivation of the gradient, the second derivative
and the gauge mode and the effect of the step rules are given in
Supplementary Notes~1 and 2.

\runin{VESPAW calculations.}
The VESPAW calculations use scalar-relativistic PAW setups, a cutoff for
the auxiliary density of twice that of the wave functions, and the PBE
functional~\cite{perdew96_prl77_3865} unless stated otherwise. The
controlled failures use MgO with the local hybrid SCAN0r, 25\% exact
exchange in the onsite terms, at the $\Gamma$ point and the default
energy tolerance $\delta=10^{-6}$~Ha. For the validation, NiO is
antiferromagnetic (AFM-II) in a rhombohedral cell with four atoms, with
PBE+$U$ ($U-J=6.3$~eV on Ni~$3d$~\cite{dudarev98_prb57_1505}) or local
hybrids with 25\% exact exchange. The spin order is prepared by a short
friction run with an orbital potential, which is then switched off, and
all optimizers start from this state. The reference is Car-Parrinello
friction dynamics~\cite{car85_prl55_2471}, which minimizes the same
functional in the space of orbitals. For bulk semiconductors, metals,
antiferromagnetic and mixed-valent oxides and local hybrid functionals,
HM with three eigensolvers and friction dynamics reach the same free
energy within 0.2~meV. The larger deviations found for direct
minimization of the oxides and for HM of a water layer in vacuum all lie
above the minimum. Systems, cutoffs, $k$-point meshes and all
validation energies are listed in Supplementary Table~1.

\runin{Descent probe C1 in VASP.}
C1 restarts the final orbitals of the tested run (\texttt{ISTART=1})
with direct minimization (\texttt{ALGO=All}), all other settings
unchanged, an energy tolerance of $10^{-7}$~eV and up to 300
iterations (600 for the $f$-electron oxides). $\mathcal A_{\mathrm{ret}}$
is the free energy of the first step of the restart, which evaluates
the returned orbitals, with the density they generate and the
occupations of the probe, before any update of the orbitals. For the
282 W-free converged entries that pass, $\mathcal A_{\mathrm{ret}}$ is
within $\delta$ of the free energy reported by the run in 248 cases and
within $10\delta$ in all of them (median $0.14\delta$), so the first
step reproduces the energy of the returned state. Larger differences
appear only for failures, where the reported energy can belong to a
mixed density (Supplementary Note~7). With the reported free energy as
reference, the result at $10\delta$ changes for 6 of the 339 entries.
$\mathcal A_{\mathrm{end}}$ is
the free energy at the end of the probe. For FeO, where \texttt{ALGO=All}
with the default step became unstable, \texttt{TIME=0.1} was used.

\runin{Janak test C2.}
We evaluate $\partial E/\partial f_n$ for the lowest unoccupied and
highest occupied states by the second-order one-sided difference
$[-3E(0)+4E(h)-E(2h)]/(2h)$ with $|h|=0.01$ electrons, requiring five
frozen-orbital energy evaluations: one at the unchanged occupations
and two for each state. Any code that returns the energy of given
orbitals at given occupations can do this. In VASP, the converged
orbitals are read with fixed occupations (\texttt{ISMEAR=-2},
\texttt{FERWE}/\texttt{FERDO}), and a single step recomputes the
density and the energy of these orbitals (\texttt{ALGO=Eigenval}). With
\texttt{ALGO=None} the stored eigenvalues are reused and the energy is
not that of the frozen orbitals. For the $f$-electron oxides, the
spin-up frontier states are tested. The MgO tests in VASP use the
$\Gamma$ point with PBE, r$^2$SCAN, SCAN, HSE06 and PBE0. NiO in VASP is
antiferromagnetic with PBE+$U$ in a cell of 64 atoms at the $\Gamma$
point with $\delta=10^{-7}$~eV, and LiMn$_2$O$_4$ uses the cubic cell
of 56 atoms with $\delta=10^{-4}$~eV.

\runin{Pass thresholds.}
The pass thresholds are deliberately conservative operational choices.
C1 fails if the probe lowers the free energy by more than $10\delta$,
where $\delta$ is the energy tolerance per cell that the run claims to
have met. For a tolerance given per atom,
$\delta=N_{\mathrm{at}}\delta_{\mathrm{at}}$. The factor is motivated,
not derived: if the energy approached its limit geometrically with a
ratio $r\leq0.9$, a last change below $\delta$ would leave an energy
error of at most $9\delta$ (Supplementary Note~7). We do not rely on
this assumption: all C1 failures reported in Table~\ref{tab:certify}
are at least $300\delta$, except four database cases at
11--$33\delta$, and increasing the threshold from $10\delta$ to
$30\delta$ would change the classification of only three. For C2, if
the Hamiltonian is the functional derivative of the implemented energy,
$\Delta_n=0$ is exact, and its numerical evaluation has a truncation
error $h^2E'''/3$, of second order in the occupation change. The error
budget of C2 in VASP (Supplementary Notes~7 and 9) has five parts. The
truncation error is below 0.2~meV: $\Delta_n$ of MgO changes by less
than this between $|h|=0.005$ and 0.02. The scatter of the
frozen-orbital energies reaches 1.5~meV for NiO with PBE+$U$, where
$\Delta_n$ varies with $|h|$ without a trend. A finer integration grid
(\texttt{ADDGRID}), a cutoff of 800~eV and a larger smearing change
$\Delta_n$ by at most 0.2~meV. With SCAN, $\Delta_n$ has a systematic
offset of 3.0 to 3.6~meV, compared with at most 1.3~meV with r$^2$SCAN
and 0.7~meV with PBE. Its origin is not established, and it limits the
resolution of C2 for SCAN. Finally, a residual of the tested state adds
to $\Delta_n$: for Ce$_2$O$_3$, $\Delta_H$ is 7.3~meV at the returned
state and 1.0~meV at the end point of C1, which is why C2 is applied
there. We call C2 failed if $|\Delta_n|>5$~meV. This threshold exceeds
the other contributions for the functionals tested, but only by
1.4~meV for SCAN, and a value between 5 and 10~meV should be confirmed
with tighter settings before it is read as an inconsistency.

\runin{$f$-electron oxides.}
UO$_2$ and Ce$_2$O$_3$ were computed with VASP~6.4.2 (\texttt{vasp\_gam}),
PBE PAW potentials, a plane-wave cutoff of 520~eV,
\texttt{PREC=Accurate}, \texttt{LASPH}, Gaussian smearing of 0.01~eV and
the $\Gamma$ point. UO$_2$: fluorite structure with $a=5.47$~\AA, 12
atoms, collinear type-1 antiferromagnet, PBE+$U$ in the form of Dudarev
et al.\ with $U=4.5$~eV and $J=0.54$~eV on U $5f$. Ce$_2$O$_3$: A-type
structure, $2\times2\times1$ cell with 20 atoms, ferromagnetic, $U=5$~eV
on Ce $4f$. A calculation with PBE is followed by one with $U/2$ and
then one with full $U$, each restarted from the orbitals of the previous
step, with Pulay mixing (\texttt{ALGO=Normal}) and up to 300
iterations. These are minimal test cells, and the energies are not
meant as physical results (Supplementary Note~7).

\runin{Database protocol.}
Structures were taken from the Materials
Project~\cite{jain13_aplmater1_011002} through its application
programming interface~\cite{ong15_cms97_209} and recomputed with its
static protocol as implemented in pymatgen~\cite{ong13_cms68_314}
(\texttt{MPStaticSet}): PBE+$U$ with the $U$ values of the
database~\cite{jain11_prb84_045115}, ferromagnetic initial moments,
520~eV cutoff, its $k$-point density, \texttt{LMAXMIX=4} (6 if the cell
contains an element heavier than Ba),
\texttt{ALGO=Fast} (Davidson followed by RMM-DIIS with Pulay mixing),
and an energy tolerance of $5\times10^{-5}$~eV per atom, that is
$\delta=N_{\mathrm{at}}\times5\times10^{-5}$~eV per cell. We changed two
settings of this input set: Gaussian smearing of 0.05~eV instead of the
tetrahedron method, so that the reported energy is the free energy
tested by C1, and a limit of 200 instead of 100 iterations. Each run
starts from scratch, from atomic densities and random orbitals, without
the output of a preceding relaxation.
For the targeted survey, each compound is represented by the database
structure of the experimentally known polymorph with the lowest energy
above the hull, in its primitive cell. These calculations do not
reproduce the energies stored in the database, which come from a
different workflow: they test this modified protocol, not the database
entries.

\runin{Census.}
The population was defined before any calculation: oxides labelled
experimentally observed (\texttt{theoretical=False}) that contain at least one of
V, Cr, Mn, Fe, Co, Ni and Cu, with a ferromagnetic, ferrimagnetic or
antiferromagnetic ground state in the database (\texttt{is\_magnetic}),
an energy above the hull of at most 25~meV/atom and at most 20 sites,
excluding the 22 compounds of the targeted survey (366 entries, database
version 2026.09.28). All entries were computed with the protocol above
in VASP 6.4.3 on 12 cores and then probed with C1 (\texttt{ALGO=All},
\texttt{EDIFF}$=10^{-7}$~eV, at most 300 iterations). The other
calculations of this work used VASP 6.4.2. The two versions agree for
CoO and NiO, with C1 giving $-789.3$~meV for CoO in both. Six
entries that stopped at the start with a symmetry error were computed
with \texttt{SYMPREC}$=10^{-4}$. Runs in which C1 reached 300 iterations
were repeated with a limit of 1000. If C1 did not converge within this
limit (Na$_3$Co$_2$SbO$_6$ among the entries without W), we use the
lowest energy along the trajectory, a lower bound on the decrease.
Runs that do not report convergence within 200 iterations are counted
separately. Four of the W-free runs that report convergence needed more
than 100 iterations and all four fail C1. With the original limit of
100 the rate at $10\delta$ would be 53 of 335 (15.8\%). All entries
are computed in the chemically primitive cell and start from the
ferromagnetic moments of the protocol, also those whose database
ordering is antiferromagnetic or ferrimagnetic. Rates are relative to the calculations that report
convergence. The POTCAR files are from PBE.54, except \texttt{W\_pv},
which is absent from it and was taken from PBE.52 (06Nov2007). The control uses \texttt{W\_sv} from PBE.54. The Wilson 95\% intervals of
Fig.~\ref{fig:prevalence}b indicate the statistical uncertainty of a
subgroup rate read as an estimate for compounds of that kind.

\runin{Controls.}
The selection rules and random seeds of the controls were fixed in a
script before any control was run. The smearing and optimizer controls
draw from the W-free failures and passes whose run and C1 converged,
that were computed without the change of \texttt{SYMPREC} and whose
$k$-point mesh has at least four points. For the smearing control, 25
failures and 25 passes (seed 2026) and CoO, FeO and NiO of the targeted
survey were computed with \texttt{ISMEAR=-5} and otherwise unchanged
settings, followed by C1 with Gaussian smearing of 0.05~eV as in the
census. For the magnetic start, all 42 W-free converged entries labelled
antiferromagnetic were computed in the database cell (larger than the
census cell for 12) with the database moments. For the optimizer
control, 20 failures (seed 2027), CoO and FeO were restarted from the
end point of C1 with Pulay mixing (\texttt{ALGO=Normal}, \texttt{EDIFF}
as in the census). An end point counts as stable if this run reports
convergence within $\max(10\delta, 1~\mathrm{meV})$ of the free energy
of C1 and within 0.1~eV of its gap. For the restart control, all 57 W-free
failures and 25 passes (seed 2028, same filter) were computed again
with the census settings, and from the returned orbitals we ran both a
restart with the same \texttt{INCAR} (\texttt{ISTART=1}) and C1. The
restart detects an entry if it ends more than $10\delta$ below the
reported free energy. It would be accepted as a replacement for C1 if
it detected at least 80\% of the entries that fail C1 in the same
calculation and flagged at most two passes. The criteria were fixed before the
runs: for the tetrahedron start, at least 15 of the 25 failures and at
most 4 of the 25 passes failing at $10\delta$. The
step dependence of C2 was tested with $|h|=0.005$, 0.01 and 0.02 on MgO with PBE, SCAN and r$^2$SCAN, for PBE and SCAN also with \texttt{ADDGRID}, a cutoff of 800~eV and smearing of 0.05~eV, and on NiO with PBE+$U$. Damped electronic dynamics (\texttt{ALGO=Damped},
\texttt{TIME}${}=0.02$, 0.05 and 0.10) and \texttt{ALGO=All} were
started from the returned orbitals of UO$_2$ (recomputed), CoO and NiO. The criterion fixed in advance, a growth rate of the residual per unit
of $n\,\texttt{TIME}$ that agrees within 30\% across the time steps,
was not met (Supplementary Note~9). All controls used VASP 6.4.3.
\end{methods}

\section*{Data availability}
The input files of all VESPAW and VASP calculations, the Materials
Project identifiers of all entries of the census, the selection and
analysis scripts and the per-system results are provided with the
\SI{} (\texttt{inputs.zip}) and will be archived at Zenodo [DOI to be
provided at the proof stage]. VASP POTCAR files are not
redistributed for licensing reasons.

\section*{Code availability}
VESPAW is distributed under the GNU General Public License, version~3,
and the version used here, 1.0.2, is archived at
Zenodo~\cite{sotoudeh26_vespaw_zenodo}. The scripts that run C1 and C2
on top of VASP and analyse the results are provided with the input
files. VASP is a commercial code.

\bibliographystyle{naturemag}
\bibliography{refs}

\begin{addendum}
 \item The author thanks Axel Gro\ss\ (Ulm University) and M. Saiful Islam (University of Oxford) for fruitful discussions. The author acknowledges financial support from the German Research Foundation (DFG) through Project ID 390874152 (EXC 2154, POLiS Cluster of Excellence). Computer time provided by the state of Baden-W\"urttemberg through bwHPC and the German Research Foundation (DFG) through grant no INST 40/575-1 FUGG (JUSTUS2 cluster) is gratefully acknowledged.
 \item[Author Contributions] M.S. conceived the work, developed the theory and the code, performed all calculations and wrote the manuscript.
 \item[Competing Interests] The author declares no competing interests.
 \item[Correspondence] Correspondence and requests for materials should be addressed to Mohsen Sotoudeh (email \href{mailto:mohsen.sotoudeh@kit.edu}{mohsen.sotoudeh@kit.edu}).
\end{addendum}

\clearpage
\includepdf[pages=-]{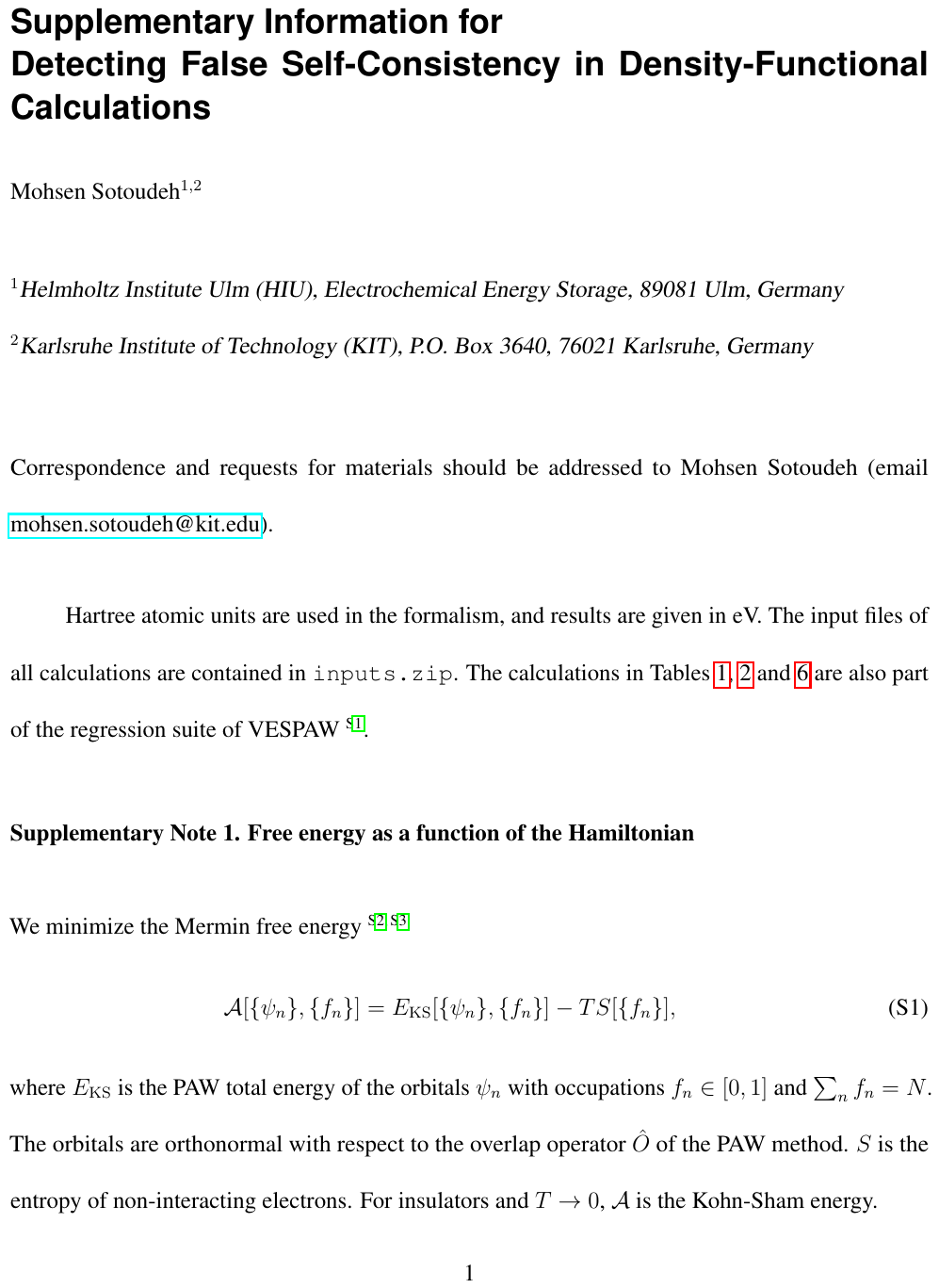}

\end{document}